# High Level Synthesis with a Dataflow Architectural Template

Shaoyi Cheng and John Wawrzynek
Department of EECS, UC Berkeley, California, USA 94720
Email: sh_cheng@berkeley.edu, johnw@eecs.berkeley.edu

*Abstract*—In this work, we present a new approach to high level synthesis (HLS), where high level functions are first mapped to an architectural template, before hardware synthesis is performed. As FPGA platforms are especially suitable for implementing streaming processing pipelines, we perform transformations on conventional high level programs where they are turned into multi-stage dataflow engines [1]. This target template naturally overlaps slow memory data accesses with computations and therefore has much better tolerance towards memory subsystem latency. Using a state-of-the-art HLS tool for the actual circuit generation, we observe up to 9x improvement in overall performance when the dataflow architectural template is used as an intermediate compilation target.

## I. INTRODUCTION

As the complexity of both the FPGA devices and their applications increase, the task of efficiently mapping the desired functionality is getting ever more challenging. To alleviate the difficulty of designing for FPGAs, there has been a trend towards using higher levels of abstraction. Tools taking in high-level function specifications and generating hardware IP blocks have been developed both in academia [2], [3] and industry [4], [5]. Of course, the semantics of the high level languages like C/C++ are vastly different than the description of hardware behavior at clock cycle granularity. The tools often try to bridge this gap by fitting the control data flow graph (CDFG) of the original program into particular hardware paradigms such as Finite State Machine with Datapath (FSMD). Depending on the nature of the application, these approaches may or may not generate hardware taking full advantage of what the FPGA has to offer. User guidance in the forms of directives or pragmas are often needed to expose parallelism of various kinds and to optimize the design. An important dimension of the space is in the mechanism with which memory data are accessed. Designers sometimes need to restructure the original code to separate out memory accesses before invoking HLS. Also, it is often desirable to convert from conventional memory accesses to a streaming model and to insert DMA engines [6]. Further enhancements can be achieved by including accelerator specific caching and burst accesses.

In this paper, we realize an intermediate architectural template (section II) that will complement existing work in HLS. It captures some of the common patterns applied in optimizing HLS generated designs. In particular, by taking advantage of the FPGAs as throughput-oriented devices, it structures the computation and data accesses into a series of coarse-grained pipeline stages, through which data flows. To target this architectural template, we have developed a tool to slice the original CDFG of the performance critical loop nests into subgraphs, connected by communication channels (section III). This decouples the scheduling of operations between different subgraphs and subsequently improves the overall throughput in the presence of data fetch stalls. Then, each of the subgraphs is fed to a conventional high-level synthesis flow, generating independent datapaths and controllers. FIFO channels are instantiated to connect the datapaths, forming the final system (section IV). The performance, when compared against directly synthesized accelerators, is far superior (section V), demonstrating the advantage of targeting the dataflow architectural template during HLS.

## II. THE DATAFLOW ARCHITECTURAL TEMPLATE

Currently, HLS tools use a simple static model for scheduling operations. Different parts of the generated hardware run in lockstep with each other, with no need for dynamic dependency checking mechanisms such as scoreboarding or load-store queueing. This rigid scheduling of operators, while producing circuit of simpler structure and smaller area, is vulnerable to stalls introduced by cache misses or variable latency operations. The entire compute engine is halted as the state machine in the controller waits for the completion of an outstanding operation. This effect becomes very pronounced when irregular offchip data accesses are encoded in the function. Under these circumstances, the traditional approach where data movements are explicitly managed using DMA may not be effective as the access pattern is not known statically. Also, there may not be sufficient on-chip memory to buffer the entirety of the involved data structure. As a result, the overall performance can deteriorate significantly.

To alleviate this problem, instead of directly synthesizing the accelerator from the original control dataflow graph, we first map the input function to an architecture resembling a dataflow engine. Figure 1 illustrates this mapping for a very simple example. The original function is broken up into a set of communicating processes, each of which can be individually turned into an accelerator. The memory subsystem is assumed to be able to take in multiple outstanding requests.

The mapping process can distribute operations in the original function into multiple stages. This process can of course generate a large number of potential implementations. At one end of the spectrum, the entire function is fit into a single stage—in which case we have a typical HLS generated accelerator, where everything is coupled with everything else through the static schedule. At the other end of the spectrum, each individual operation can be scheduled independently as a standalone module, getting activated whenever its inputs are available—we essentially have a fine grained dataflow machine with very high area overhead. The optimal design points lie somewhere in between these two extremes. In section III, we present one partitioning algorithm which aims to localize data fetch stalls, and thus addresses the main factor for the performance degradation mentioned earlier. It by no means covers the entire space of different partitioning granularity, but should have reaped most of the benefits in mapping to this intermediate dataflow architecture.

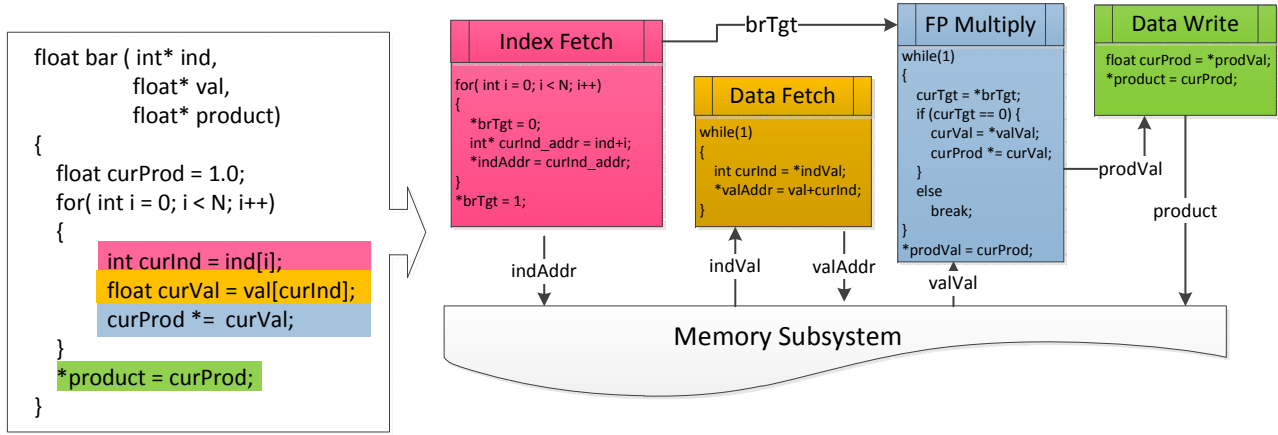

Fig. 1. Conversion to a dataflow pipeline

For our example, the change in overall execution schedule can be seen in figure 2. The occasional off-chip data fetches no longer affect the performance of the processing engine as the long latency floating point multiply shadows the stalls introduced by these cache misses. As long as the overall bandwidth provided by the memory subsystem can satisfy the computation, the latency of memory accesses can be tolerated. This overlapping of computation and communication is naturally provided by the architecture, and provides significant boost in performance, as will be shown later.

## III. MAPPING TO THE DATAFLOW ARCHITECTURAL TEMPLATE

Mapping from a standard CDFG to the dataflow architectural template involves a partitioning process where dependency edges are cut and nodes are assigned to different stages in the pipeline template. To maximize the performance of the final circuit, several factors need to be considered during this partitioning process. First, circular dependencies between nodes need to be contained within stages. These strongly connected components (SCCs) in CDFG are associated with loop carried dependencies, and are the limiting factors for how aggressively loop iterations can be overlapped. The initiation interval (II) of loops are dictated by the latency of these cycles. As the communication channels will always add latency, having parts of a SCC distributed across different stages would increase the II of the iterations, as they are now executed in a distributed manner. The same observation was made in [7], albeit in a different context. Secondly, as we have shown in section II, with memory operations separated from dependency cycles involving long latency computation, we can have cache misses completely hidden by the slow rate of data consumption. Thirdly, to localize the effects of stalls introduced by cache misses, the number of memory operations in each stage should be minimized, especially when they address different parts of the memory space.

### A. The Partitioning Algorithm

In Algorithm 1, the steps taken to achieve the aforementioned requirements are detailed. The SCCs are collapsed into new nodes, which together with the original nodes in the CDFG, are topologically sorted. The obtained directed acyclic graph is traversed and a new pipeline stage in the template is created whenever a memory operation or an SCC with long latency computation is encountered. Here, long latency operations are those which cannot be completed within one clock cycle. The details of the hardware device and the design's timing constraints determine which operations are long latency. In our experiments, we leverage Xilinx's Vivado HLS to generate the these timing numbers. With a target clock frequency of 150MHz, for instance, a 32 bit integer addition can be completed within a one clock cycle while a floating point multiply takes four cycles. These numbers are accurate as we eventually use Vivado HLS for our HDL generation.

The semantics of the input high level language often create dependencies implicitly carried by memory accesses. Before the partitioning algorithm is run, explicit edges between memory access operations are added to ensure these dependencies are not violated. This operation essentially divides up the memory space into disjoint regions. User annotations are used to provide guidance to the tool as alias analysis may produce over conservative results when pointer arithmetic or irregular memory accesses are present in the original code.

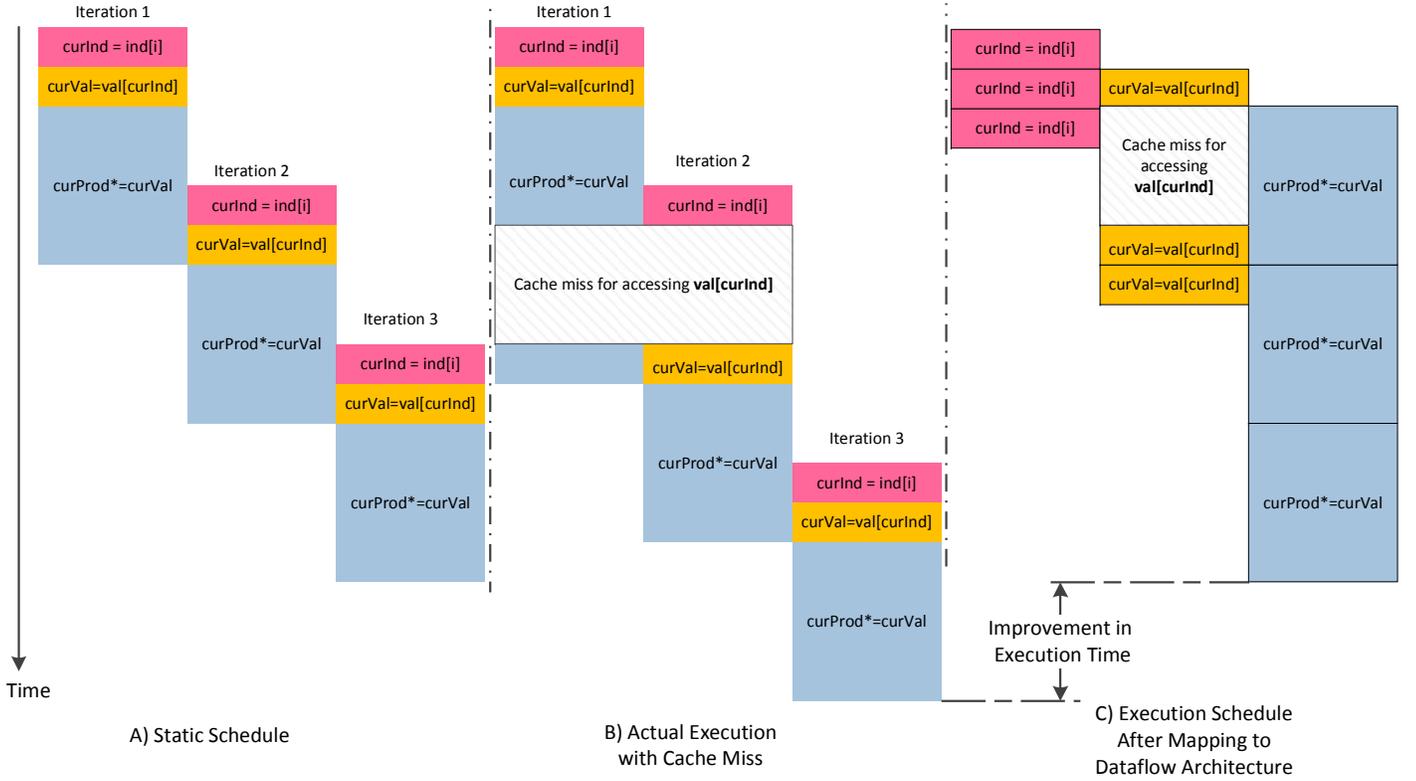

Fig. 2. Execution schedule of accelerator with and without mapping to dataflow architecture

**Algorithm 1** Partitioning algorithm
1: **procedure** PARTITIONCDFG($G$)
2:    $SCCs \leftarrow allStronglyConnComps(G)$
3:    $DAG \leftarrow collapse(SCCs, G)$
4:    $TopoSortedNodes \leftarrow topologicalSort(DAG)$
5:    $LongSCCs \leftarrow getSCCWithLongOp(SCCs)$
6:    $MemNodes \leftarrow findLdStNodes(G)$
7:    $MemLongSCC \leftarrow LongSCCs \cup MemNodes$
8:    $allStages \leftarrow \{\}$
9:    $curStage \leftarrow \{\}$
10:    **while** $TopoSortedNodes \neq \emptyset$ **do**
11:       $curNode \leftarrow TopoSortedNodes.pop()$
12:       $curStage \leftarrow curStage \cup curNode$
13:       **if** $curNode \in MemLongSCC$ **then**
14:          $allStages \leftarrow allStages \cup curStage$
15:          $curStage \leftarrow \{\}$
16:       **end if**
17:    **end while**
18:    **return** $allStages$
19: **end procedure**

After the partitioning, the edges spanning two stages will be converted to communication channels for data/control tokens to be passed between them. From each stage's perspective, these are just normal load/store operations from/to special pointers corresponding to communication interfaces. During the actual hardware generation, FIFOs will be instantiated and connected.

### B. Potential Optimizations

*1) Trade-off Between Computation and Communication:* As the decoupled processing modules are eventually connected together to form a pipeline, each pair of communication primitives inserted requires an instantiation of a FIFO. On FPGAs, the area costs for FIFOs, even ones with a relatively small number of entries, can be significant. Consequently it is often better to duplicate some computation rather than creating a new hardware queue between two modules. Currently, we do not duplicate long latency computation (e.g. multiply) or memory accesses as these are the operations we aimed to separate during the partitioning step. However, some of the most frequently occuring SCCs, i.e. increment of loop counters, still provide opportunities for this optimization.

*2) Memory Optimization:* As mentioned in section III-A, after a memory operation is added to a stage, a new stage is created. All the consumers of the data requested are thus decoupled from the stage issuing the memory request. This transformation allows for many outstanding requests to be pipelined into the memory subsystem. For accessing consecutive memory locations, these operations are automatically converted to burst accesses, improving efficiency of memory bandwidth usage.

In our flow, the partitioning of the memory space has also provided an opportunity to better customize the hardware for memory access on the FPGA fabric. Each independent data access interface, corresponding to one memory partition, can be supported differently according to the nature of the address stream it generates. For instance, when streaming type loads are concatenated into bursted accesses, they would not benefit from having on-chip buffer due to the absence of data reuse, while random memory accesses can be helped with a general purpose cache, whose size and associativity can be tuned according to the runtime profile.

## IV. Hardware Generation

With the high level function mapped to the dataflow architecture, we then leverage conventional HLS tools for the RTL generation. The mapping step described in section III is based on the LLVM infrastructure [8]. The LLVM front end parses C/C++ functions and convert them into the single static assignment (SSA) form, that facilitates dependency tracking between operations, making the implementation of the algorithm easier. To provide HLS tools with a valid input, we convert the LLVM intermediate representation for each stage in the pipeline to synthesizable C, with a complementary TCL script to drive the tool and connect the stages with FIFOs.

The approach we take in converting LLVM IR to C is rather simplistic. Instead of recovering the higher level semantics, we simply generate a one to one mapping of each LLVM instruction to C statement. The only common llvm instruction which does not have a simple equivalence in C is the $\phi$ operation, which assigns values based on the predecessor of the current basic block. Our flow creates a different assignment statement in each of the predecessor basic blocks as specified by the $\phi$ operation. The $\phi$ operation itself is then removed.

The steps involved, i.e., mapping of the original input function to the architecture template and the partitioning and hardware generation, are summarized in figure 3.

## V. Experimental Evaluation

To evaluate the benefits of our approach, we took several benchmark kernels and processed them with our flow. For sparse matrix vector (SpMV) multiply—our first kernel—compressed sparse row (CSR) format is used to store the matrix, where the loads of the floating point numbers to be multiplied depend on the data in an index array. The next two kernels, Knapsack and Floyd-Warshall, are both dynamic programming problems. The memory addresses to be accessed are derived from the results of computation. Our final kernel, Depth first search (DFS), is a widely used graph algorithm. The benchmarked version operates on pointer based data structure and uses a stack. All these kernels are sequential code with non-regular control flow and memory access patterns. The input dataset for each benchmark, as described by Table I, is also too big to fit entirely on chip. Due to the presence of statically unknown off-chip memory accesses, these benchmarks are good cases where our algorithm can play a significant role in improving the overall performance. On the other hand, for problems where the entire data set can be buffered on chip or when the memory access patterns are regular, our approach would offer little performance advantage over the conventional DMA+accelerator approach [6].

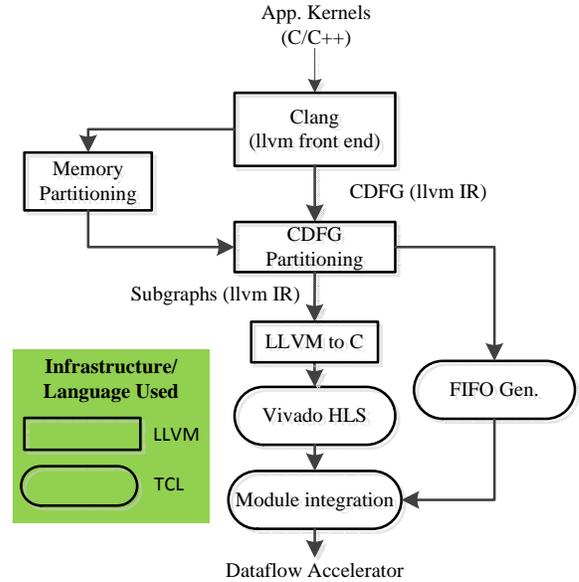

Fig. 3. Pipeline Generation Flow

TABLE I
INPUT DATA SET FOR THE BENCHMARKS

| Benchmark | Description of Input Data | Size of Input Data |
|---|---|---|
| SpMV Multiply | Matrix dimension = 4096<br>Density of Matrix = 0.25 | ≈ 16 MB |
| Knapsack | Weight Limit = 3200<br>Number of Items = 200 | ≈ 5 MB |
| Floyd-Warshall | Number of Nodes = 1024 | ≈ 8 MB |
| Depth-First Search | Number of Nodes = 4000<br>Number of Neighbors per Node = 200 | ≈ 3 MB |

The FPGA device used for our evaluation is the Zynq-7000 XC7Z020 FPGA SoC from Xilinx, installed on the ZedBoard evaluation platform. The SoC contains two parts: an ARM-processor based processing system (PS), and the programmable logic (PL). The baseline for our evaluation is the performance of each software kernel running on the ARM core in the PS. The core is an out-of-order, dual-issue hard processor running at 667MHz. The Zynq platform also provides two options for the accelerators in PL to access the main memory subsystem: through the accelerator coherence port (ACP), or the high performance (HP) port. The former connects to the snoop control unit in the processing system and thus uses/modifies the processing system's on chip cache. The HP port connects directly to the memory controller, which necessitates the flushing of cache lines by the processor if cached memory is accessed by the accelerator. In either case, if memory states are also buffered in the PL with caches, they need to be explicitly pushed to the processing system side after the accelerator finishes running. As both the ACP and the HP are slave ports, they provide no mechanisms to extract data from the programmable logic when the ARM processor is running. The interaction between the generated accelerators and the main pieces of the FPGA SoC is shown in figure 4.

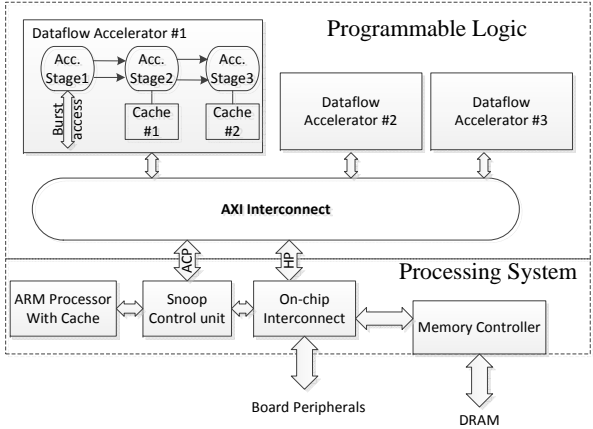

Fig. 4. Implementation of Dataflow Accelerators in FPGA SoC

In our study, Vivado HLS, a state-of-the-art high level synthesis tool provided by Xilinx, is used for generating the conventional accelerator as well as the individual stages in the dataflow accelerators generated by our flow. With the target clock period set to 8ns during HLS, the tightest timing constraints post place & route implementations managed to meet range from 111 to 150MHz. All design points shown in this section use the highest achievable frequency as the actual operating clock frequency.

### A. Performance Comparisons

In figure 5, performance of the different implementations are presented. Conventional accelerators and dataflow accelerators with different memory subsystem configurations are compared. All the numbers are normalized to the baseline.

In all four benchmarks, conventional accelerators directly generated from software kernels using the HLS flow actually have lower performance than the hard processor. Even with on-PL caches, these accelerators only manage to achieve throughput less than 50% that of the baseline. The superscalar, out-of-order ARM core is capable of exploiting instruction level parallelism to a good extent and also has a high performance on-chip cache. The additional parallelism extracted by the HLS tool is evidently not enough to compensate for the clock frequency advantage the hard processor core has over the programmable logic and the longer data access latency from the reconfigurable array.

With our methodology, the processing pipelines generated are rather competitive against the hard processor, even without a reconfigurable cache. For SpMV multiply, knapsack and Floyd-Warshall, when the dataflow accelerators are directly connected to the PS through the ACP, the average performance is 2.3x that of the baseline—representing an 8.4x gain over the conventional accelerators. Upon the addition of caches, the average runtime of the dataflow accelerators was reduced by 18.7%, while that of the conventional accelerators was cut by 45.4%. The gap between their performance is thereby reduced from 8.4 to 5.6 times. This difference in improvement is due to conventional accelerators' sensitivity to the latency of data accesses, which is also manifested by its performance degradation of 40% when the uncached HP port is used instead of ACP.

It is also apparent that our approach has its limitations, as demonstrated by its ineffectiveness in the depth first search benchmark. The kernel performs very little computing but lots of memory accesses. The use of a stack in DFS also creates a dependence cycle through the memory and consequently, the performance is fundamentally limited by the latency of memory access. Thus there were only small differences between the performance of the conventional accelerator and the dataflow accelerator. Besides, the memory access pattern does not provide many opportunities for optimizations. As a result, both accelerators achieves performance far below that of the baseline.

Overall, for kernels suitable for FPGA acceleration, there is a significant performance advantage in using an intermediate dataflow architectural template. If we compare the best results achieved with the dataflow accelerators to that of the conventional accelerators, we see improvement of 3.3 to 9.1 times, with an average of 5.6.

### B. Area comparison

To quantify the impact of our proposed methodology on area, we have compared the FPGA resouce usage of conventional accelerators and the dataflow accelerators. Table II shows the results, where each acclerator is complemented with two different memory subsystem configurations.

TABLE II
RESOURCE USAGE OF ACCELERATORS.

| Benchmark | | ACP | | | ACP + 64KB Cache | | |
|---|---|---|---|---|---|---|---|
| | | LUT | FFs | BRAM | LUT | FFs | BRAM |
| SpMV Multiply | Con.Acc | 9873 | 9116 | 10 | 7918 | 6792 | 21 |
| | Dataflow Acc | 8577 | 8837 | 10 | 6718 | 6788 | 21 |
| | % change | -13.1 | -3.1 | 0 | -15.2 | -0.1 | 0 |
| Knapsack | Con.Acc | 7672 | 7490 | 8 | 6573 | 5885 | 21 |
| | Dataflow Acc | 8089 | 8787 | 8 | 6970 | 7256 | 21 |
| | % change | +5.4 | +17.3 | 0 | +6.0 | +23.3 | 0 |
| Floyd-Warshall | Con.Acc | 2491 | 3528 | 0 | 3806 | 4629 | 19 |
| | Dataflow Acc | 7659 | 7210 | 0 | 8995 | 8309 | 19 |
| | % change | +207.5 | +104.3 | 0 | +104.4 | +79.5 | 0 |
| DFS | Con.Acc | 4810 | 4929 | 4 | 4931 | 4594 | 21 |
| | Dataflow Acc | 8509 | 7813 | 4 | 7436 | 6298 | 21 |
| | % change | +76.9 | +58.5 | 0 | +50.8 | +37.1 | 0 |

The difference in area between the dataflow accelerators and the conventional accelerators is effected by two factors. When the dataflow architectural template is used, there are always additional costs associated with the communication channels. Meanwhile, as the original programs are partitioned into subgraphs and separately turned into hardware, the depth

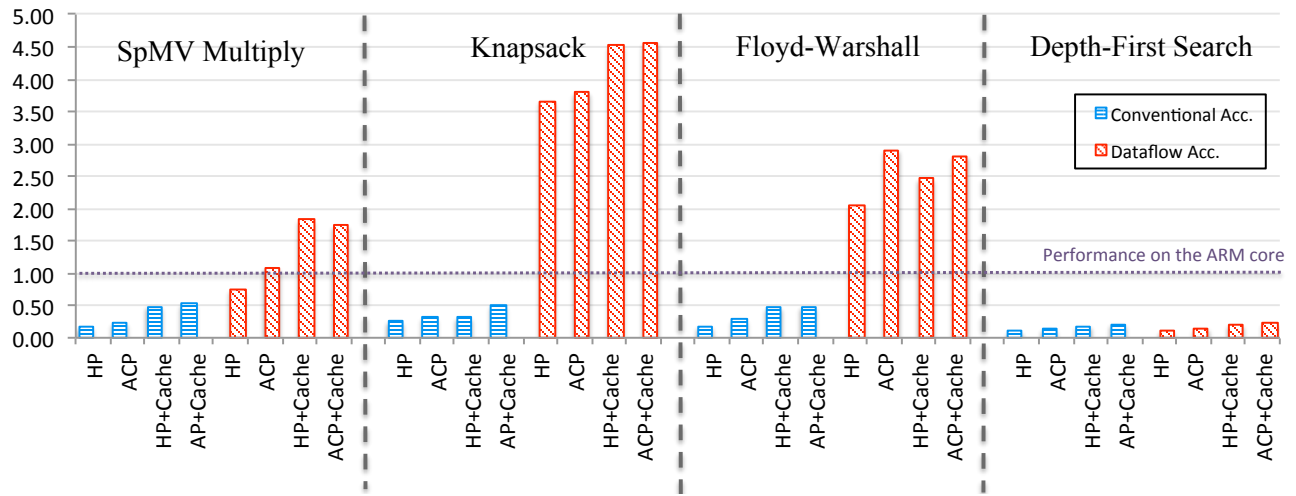

Fig. 5. Performance of Conventional Accelerators and Dataflow Accelerators, normalized to the hard ARM core. Implementations with caches use Xilinx System Cache IP, configured to be 64KB and 2-way associative

of the internal pipeline in each module is reduced, resulting in area savings. The overall change therefore depends on which factor plays a larger role, and is ultimately application specific.

## VI. CONCLUSION

This paper presents a new architectural template that can be used as an intermediate target for high level synthesis. The accelerators generated using our method run significantly faster than conventional accelerators as computation and communication are automatically overlapped. Overall, our new approach produces hardware engines with an average 5.6 times perfomance advantage over normal HLS with no intermediate architecture mapping.

## ACKNOWLEDGEMENT

This research is supported by the Berkeley Wireless Research Center and the UC Berkeley ASPIRE Lab. The ASPIRE Lab is funded by DARPA Award Number HR0011-12-2-0016, the Center for Future Architecture Research, a member of STARnet, a Semiconductor Research Corporation program sponsored by MARCO and DARPA, and industrial sponsors and affiliates: Intel, Google, Huawei, Nokia, NVIDIA, Oracle, and Samsung. Any opinions, findings, conclusions, or recommendations in this paper are solely those of the authors and does not necessarily reflect the position or the policy of the sponsors.